\documentclass[12pt]{article} 
\usepackage{graphicx}
\title{Design of a Plasma Accelerator Based on Effects of Energy/Momentum Exchange in Crossed Magnetic Fields}
\author{A. R. Karimov \\
Institute for High Temperatures RAS, \\ Izhorskaya 13/19, Moscow 127412, Russia \\ Department of Electrophysical Facilities, \\ National Research Nuclear University MEPhI, \\ Kashirskoye shosse 31, Moscow, 115409, Russia\\
P. Murad \\
Morningstar Applied Physics, LLC \\ Vienna, VA 22182 USA}
\date{ }
\begin{document} 
\maketitle
\newcounter{graf}
\begin{abstract}
The creation and design of a plasma thruster are considered using a rotating cylindrical plasma flow in crossed magnetic fields. The thruster scheme uses this plasma guided by a permanent, radial and a time-dependent axial magnetic field. Namely, the acceleration effect is formed by the azimuthal electric field which comes about as a variation of an external, axial magnetic flux. Besides, it is possible to go beyond the approach of quasi-neutrality. These processes can lead to the generation of a local azimuthal time-dependent electric field which may also accelerate the rotation of the electron and ion fluids in different directions. Furthermore, estimations of this device and the plasma flow demonstrate additional thrust effects as well as can consume interplanetary medium as a fuel for driving plasma thrust.
\end{abstract}

\maketitle

\section{Introduction}

Plasma thrusters have an undeniable advantage over the conventional chemical jet or rocket engines. This follows directly from the Tsiolkovsky formula:
\begin{equation}
V_f= V_{ex}\ln\left({M_0 \over M_f}\right)\/,
\label{1_ccpt}
\end{equation}
where $V_f$ is the final velocity of the rocket, $M_0$ is the initial total mass, including propellant, $M_f$ is the final total mass of spacecraft without propellant, and $V_{ex}$ is the effective exhaust velocity. As is seen from Eq. (\ref{1_ccpt}), it is unprofitable to accelerate the propulsion apparatus owing to the ejected and total mass fraction. In order to have a large mass fraction for the delivered payload, the specific impulse of the thruster must be on the order of the total $V_{ex}$. However, the effective exhaust velocity of the jet engine is limited by the enthalpy of the used fuel. Such a limit is absent in plasma thrusters that make it possible to obtain the high-velocity exhaust velocities in a more efficient approach.

In this case, the thrust generated by electric propulsion systems is only limited by the amount of electric power that can be supplied \cite{sb,ms}. However, all these devices are characterized by a low specific impulse that comes about because of a low density of neutralizing the plasma flow. An increase in specific impulse increases the mass efficiency while also decreasing the thrust thereby increasing trip travel time. Therefore, one option of the problem is to increase the density of the plasma flow at a higher velocity.

In the present piece, we shall focus on the plasma thrusters with the crossed fields accelerating plasma ions. As a rule, in such devices (so called Hall thrusters) would use the strong magnetic field perpendicular to the flow and the axial electric field to accelerate an ion component. The magnetic field impedes the counterflow of electrons in the accelerating field which does away with the space-charge limitation. This restricts the flow and thrust of ion engines (i.e. acceleration of ion flow in this region come about in quasi-neutral conditions and there is no limitation brought about by the space charge). This feature makes the Hall thrusters being one of the most efficient types of plasma devices among plasma accelerators. For comparison, the efficiency of an ionic electrostatic accelerator such as gridded ion thrusters is limited by the Langmuir law. However, it should be borne in mind that although the Hall thrusters are mechanically simpler, physics of Hall thrusters are more complex and they achieve slightly less electric efficiency and specific impulse than ion thrusters. So, as a first step, it will be useful to consider the basic physical features of the existing Hall thrusters.

\section{The operational principles of Hall thrusters}

In Hall thrusters, the orthogonal electric $\mathbf{E}$ and magnetic $\mathbf{B}$ fields are used to produce thrust by the ion component created in the plasma (see \cite{ms,gk}). As we have said, an electron component is assumed to be fixed in the axial direction. A quasi-equilibrium situation may exist in a pressure-less plasma possessing high electron conductivity $\sigma$ when it is possible to allow  $\sigma \to \infty$ (i.e. the frictional force between electrons and ions is negligible and the corresponding term in the equation of motion for electrons and ions is absent). The plasma parameters in the acceleration zone are defined by these equations:
\begin{equation}
\partial_t \mathbf{v}_i + (\mathbf{v}_i \cdot \nabla) \mathbf{v}_i = {e \over m_i} {\Bigl (}\mathbf{E} + {1\over c}\mathbf{v}_i \times \mathbf{B}{\Bigr )}\/,  
\label{2_ccpt}
\end{equation}
\begin{equation}
\mathbf{E} = -{1\over c}\mathbf{v}_e \times \mathbf{B}\/, 
\label{3_ccpt}
\end{equation}
where $m_i$ is the ion mass, $\mathbf{v}_i$ is the ion fluid velocity, $\mathbf{v}_e$ is the electron fluid velocity, and c is the velocity of light. Here $\mathbf{E}$ and $\mathbf{B}$ represent a superposition of external and internal electric and magnetic fields, respectively. As is seen from relation (\ref{3_ccpt}), the electric field is able to accelerate ions that exist in the plasma only do to $\mathbf{B}$ when $\mathbf{v}_e\ne \mathbf{v}_i$. In this case, the electron density $ n_e$ is approximately equal to the ion density $n_i$. The external magnetic field is assumed to be large enough to magnetize electrons but not to impact ions \cite{ms,gk}:
\begin{equation}
r_{Le}\ll L \ll r_{Li}\/,
\label{L_ccpt}
\end{equation}
where $r_{Le}$ and $r_{Li}$ are the Larmor radius for electrons and ions, respectively, and $L$ is a length of the self-consistent accelerating layer.  If these requirements are fulfilled, and we can always find such external magnetic field to satisfy (\ref{L_ccpt}), then the space charge of the ions is compensated by a small mobility of electrons across the magnetic field. As a result, the electrons can experience the rotation in the azimuthal direction with a drift velocity $\mathbf{V}_E=({\bf E}\times {\bf B} / B^2)c$, and ions may move only in the axial direction. Thus, a radial magnetic field confines electrons which compensate the space charge of the flow, while an axial electric field accelerates ions.

\begin{figure}
\centering
\includegraphics[width=3.5in]{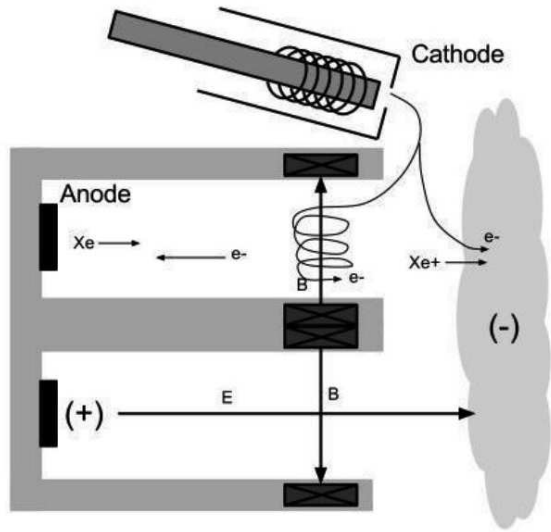}
\caption{Schematic of stationary plasma thruster \cite{gp}}
\label{f_spt}
\end{figure}

This idea is for stationary plasma thrusters where the thruster includes an anode layer in which the electrical discharge has an ${\bf E}\times {\bf B}$ configuration where the external magnetic field is radial and perpendicular to the axial electric field (see for example, \cite{ms}-\cite{kim}). A schematic of this thruster is shown in Fig. \ref{f_spt}. The size of the acceleration zone in the axial direction of stationary plasma thrusters is acceptable to be greater in thrust performance than the one in thrusters with an anode layer. Nevertheless, these thrusters are in accordance with the principle of this operation and the parameters achieved. In fact, the electron fluid in these thrusters acts as an immobile (in the axial direction) neutral medium so long as condition Eq. (\ref{L_ccpt}) is satisfied.  It should be noted that the effect of the ion acceleration process is limited by the charge polarization owing to the generation of an internal electric field having the order $\sim n_e\lambda_D$, where $\lambda_D$ is the  Debye length.

One may try to go to the limits of the magnetized electrons approximation if both ion and the electron flows move in one or the same direction.  Such acceleration may occur if plasma flow is in the crossed magnetic fields varied with time and space. Typically, this idea is similar to an end-Hall thruster and a cylindrical type Hall thruster where the acceleration is in the axial direction. The generation of an electric field brings about the interaction between the electron flow and a magnetic field (for example \cite{srf,erf}). At the same time, this system provides thrust without the need for additional current and space charge neutralization. It is also worth noting that the similar mechanism of acceleration occurs with a coaxial plasma accelerator \cite{koz08,koz09} or a magnetoplasmadynamic thruster \cite{jc}.

\begin{figure}
\centering
\includegraphics[width=3.5in]{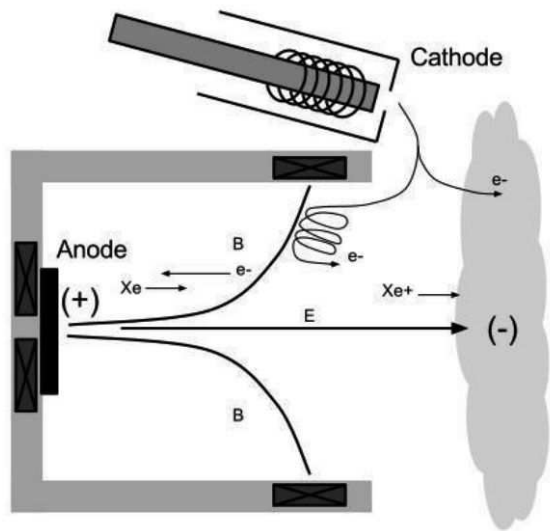}
\caption{Schematic of cylindrical Hall thruster \cite{gp}}
\label{f_CHT}
\end{figure}

The typical principle of cylindrical Hall thruster is depicted schematically in \ref{f_CHT}.
This thruster distinguishes from conventional annular and end-Hall thrusters by using a cylindrical configuration with an enhanced radial component using a cusp-type magnetic field. Such a magnetic field configuration can be created with the help of electromagnet coils \cite{rf_01,srf_07} or permanent magnets \cite{rmf_10}. Unlike conventional Hall thrusters, cylindrical hall thrusters do not have the center stem that provides a radial magnetic field. Similar to the conventional annular type Hall thruster, there is a closed ${\bf E}\times {\bf B}$ electron drift and acceleration of non-magnetized ions by the electric field defined by Eq. (\ref{3_ccpt}).  Here the radial component of the magnetic field crossed with the azimuthal electron current produces thrust. At the same time, the electrons are not confined to an axial position, and they bounce over an axial region.

\begin{figure}
\centering
\includegraphics[width=3.5in]{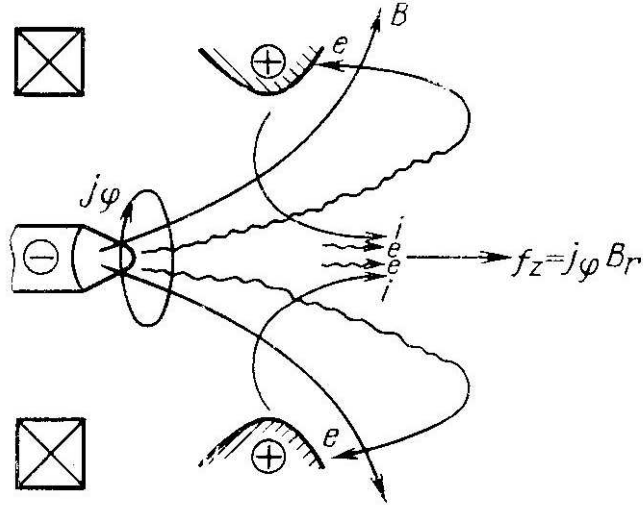}
\caption{Schematic of end-Hall thruster \cite{glk}. Here ${\bf j}_ {\varphi}$ is the sum of electron ${\bf j}_{e \varphi}$  and ion ${\bf j}_{i \varphi}$ current fluxes}
\label{f_EHT}
\end{figure}

This can be considered as an electrostatic viewer on the acceleration mechanism from an electromagnetic point of view. There is another interpretation. The azimuthal Hall current interacts with an applied radial magnetic field and accelerates the plasma axially through the Lorentz force. Such a mechanism arises naturally in the end-Hall thruster (see Fig. \ref{f_EHT}). Unlike stationary plasma thrusters and cylindrical Hall thrusters, there is no neutralizer in this scheme but the size of the acceleration zone is extremely small and is defined by the configuration of the magnetic field at the output of the magnetic coil (see \cite{rmf_10}). Thus with this appearance, the magnetic coils in these devices create a configuration where there is both a radial and axial component of the magnetic field.  

\section{ Design of plasma accelerator}

\begin{figure}
\centering
\includegraphics[width=3.5in]{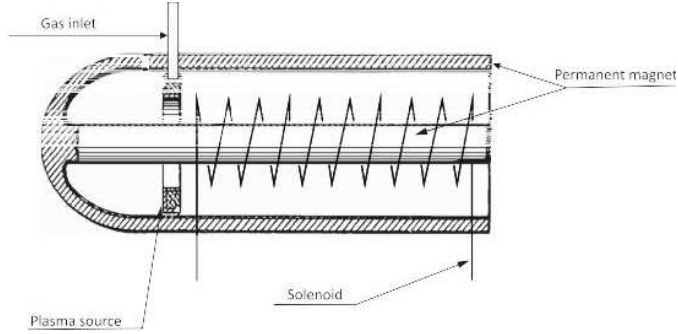}
\caption{Design of device on the base \cite{mgs}.}
\label{ACI}
\end{figure}

This acceleration mechanism for the rotating plasma flow in the external magnetic fields can be strengthened in the schematic of a thruster concept sketched in Fig. \ref{ACI}. 
This is based on the design proposed in \cite{mgs}. Here, a gas fuel is injected into the system through the gas inlet where the gas is ionized by a plasma source. The length of the discharging part of the inlet must be designed to minimize the ionization mean free path localizing the ionization of the working gas into the inlet volume. It is also worth noting that in principle, as a gas fuel, one may use the existing ambient space particles.

\begin{figure}
\centering
\includegraphics[width=3.5in]{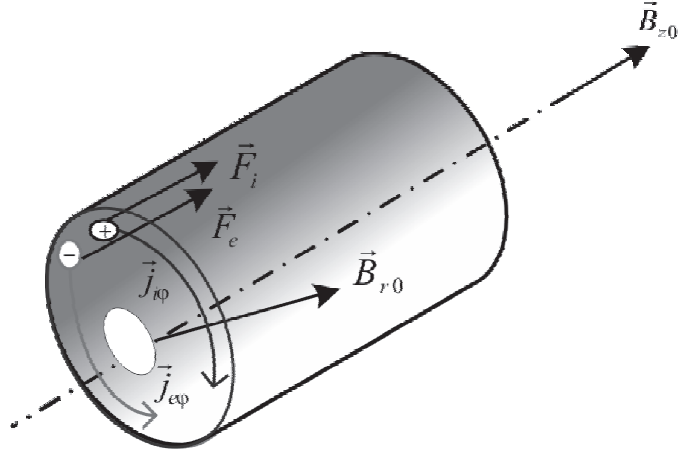}
\caption{Schematic of electron and ion acceleration region.}
\label{AC}
\end{figure}

Here we assume the permanent magnet system creates a constant radial magnetic field and the solenoid fed through a low-frequency alternating current. As a result, the configuration of the crossed magnetic fields presented in Fig. \ref{AC} uses a combination of external fields to produce a spatially homogeneous magnetic field:  
\begin{equation}
\mathbf{B}^0= B_{r0}\mathbf{e}_r + B_{z0}(t)\mathbf{e}_z\/,
\label{B_ccpt}
\end{equation}
where $B_{r0}=$const. and $B_{z0}=B_{z0}(t)$  is some known function of time. Here we assume that this field remains spatially homogeneous in plasma that is valid if:    
\begin{equation} 
\beta = {\sum _s n_s T_s \over B_0^2/ 8 \pi}  \ll 1\/,
\label{17_saxa} 
\end{equation}
where $T_e$ is the electron temperature, and $T_i$ is the ion temperature (i.e. the diamagnetic effect is considered negligibly small).

From an equation of induction written in integral form for this field $\mathbf{B}^0$
$$\int_{\gamma} \mathbf{E}^0d\mathbf{l}= - \partial_t \left( \int_{S_{\gamma}}\mathbf{B}^0 d\mathbf{S}\right)$$
it follows that the external electric field has only an azimuthal component:
\begin{equation}
E_{\varphi 0}= -{r \over 2 } \partial_t B_{z 0}\/,
\label{7_ac}
\end{equation}
which rotates electron ${\bf j}_{e \varphi}$ and ion ${\bf j}_{i \varphi}$ current fluxes in different directions. The azimuthal electron and ion flows interact with the radial magnetic $B_{r0}$   via the Lorentz force. This leads to an axial acceleration of electrons and ions in the same axial direction since the resulting forces ${\bf j}_{e \varphi}\times {\bf B}_{r0}$ and ion ${\bf j}_{i \varphi}\times {\bf B}_{r0}$ are also directed in the same direction. Owing to this process, the transformation of azimuthal momentum transfers into axial momentum. As a result, the electron and ion flows are accelerated in one axial direction sketched in Fig. \ref{f_EHT}. Thus, one can accelerate the electron and ion flows in the axial direction by accelerating them in the azimuth direction by a vortex electric field that may be generated by varying the axial magnetic field in time (the discussion of this idea was presented in \cite{km17, km_17}).

However, as is seen from Eq. (\ref{7_ac}) for a harmonic function, $B_{z 0}=B_{z 0}(t)$ creates the acceleration which can occur only for a quarter period when $\partial_t B_{z 0} \le 0$.  It implies that we have to use some electronic unit to form the signal $B_{z 0}(t)$ waveform with the required phase.

Although this point is purely a technical problem that depends on the amplitude and frequency of the original signal, we show that the problem is solvable in principle. Let the current creating the axial magnetic flux in the solenoid be:
\[\ I_0(t)=\left\{
\begin{array}{rl}
\cos\left({\pi \over T}t\right), & 0 <t < T/2, \\
0, & T/2< t<T,
\label{I_ccpt}
\end{array}\right.\]
where $T$  is the constant defining the signal period.  It is obvious that in this case, the magnetic field $B_{z 0}(t)$ will repeat the form of the signal $I_0(t)$.  On the other hand, we can present $I_0(t)$ in a formal Fourier series as:
\begin{equation}
I_0(t)={1 \over 2} \cos\left({\pi  \over T} t\right)+ {2 \over  \pi}\sum_{l=1}^{\infty}{(-1)^{l+1} \over 4l^2 - 1}\cos\left({2\pi l \over T}t\right)\/.
\label{I0_ccpt}
\end{equation}
As is seen from Eq. (\ref{I0_ccpt}), for $l>1$ the amplitude of the upper harmonics is negligible.  Therefore, we may describe the signal $I_0(t)$ with just a few lower harmonics. For example, if we are restricted by the rude case when $l=1$ and we may use the bifilar coil with a parallel winding of wires, to make the rude form of the current $I_0(t)$ in this coil. Certainly, the waveform can be approximated to the form of $I_0(t)$ if we use trifilar, tetraphilic, pentafilar, etc. coils which are able to pass higher harmonics. However, this issue is not considered in the framework of the present paper.

Now, we discuss the conditions under which this acceleration scheme can be realized. Instead of Eq. (\ref{L_ccpt}), here we must put:
\begin{equation}
a<r_{Le}< r_{Li}<b\/,
\label{L2_ccpt}
\end{equation}
Where $a$ and $b$ are the internal and external radius of the magnetic system. If we take the value $v_{e}= (T_e/m_e)^{1/2}$ as a characteristic velocity of electron flow, and $v_{i}= (T_e/m_i)^{1/2}$ as a characteristic velocity of the ion flow, then we can rewrite the condition of Eq. (\ref{L2_ccpt}) in the following form:
\begin{equation}
a< {(T_e m_e)^{1/2}\over e B_{z0}}<{(T_e m_i)^{1/2}\over e B_{z0}}<b\/.
\label{L3_ccpt}
\end{equation}
Our model presumes the absence of collisions between particles of the plasma medium. Such a situation occurs if:
\begin{equation}
\lambda \gg b\/, 
\label{4_ccpt}
\end{equation}
where $\lambda=1/(\sigma n_e)$  is the mean free path of the plasma flow, and $\sigma$  is the effective cross-section. As a characteristic estimate, we take $\sigma = \pi a_0^2$, where $a_0$ is the Bohr radius. Taking into account this value of $\sigma$, we arrive at:
\begin{equation}
b n_e \ll 10^{16}\/. 
\label{26b_ac}
\end{equation}
Moreover, from (\ref{L2_ccpt}) and (\ref{4_ccpt}), it follows that the cyclotron frequency $\omega_{sc}$ for the s-component exceeds the collision frequency $\nu_{coll}$ between the charged particle of the s-kind with particles of other kinds (i.e. the Hall parameter $h_s=\omega_{sc}/ \nu_{coll} \gg 1$) and the medium is considered to be a magnetized plasma. In this case, all transport properties of plasma are defined by the magnetic field and the medium becomes anisotropic.

We give up the generation of the magnetic field of the plasma flow (i.e. the magnetic field brought about by the movement of plasma electrons and ions) that would reduce this effect. It implies that we can neglect and ignore the spatial dependence of this magnetic field. According to the estimation \cite{km17}, such an assumption is valid for:
\begin{equation}
b^2 n_e \gg 10^{13}\/. 
\label{26a_ac}
\end{equation}
However, even with these parameters, they are beyond the range of condition in Eq. (\ref{26a_ac}) and one can expect to result in some additional thrust effects.

\section{Speculating prospects}

There may be a lot of technical implementations for this idea. Here we discussed the schematic based on the geometry proposed in \cite{mgs}. However, instead of an electromagnet here, we propose to use a permanent magnet to create the constant radial magnetic field (see Fig. \ref{ACI}). Also, we don't discuss the electronic method for the formation of signal $B_{z0}(t)$ of the required form. One can say that this is a solvable issue from the technical point of view but the solution depends on both the amplitude and frequency of the original signal. Furthermore, this problem originates from the framework of the present paper and we hope to discuss this question later.  

Here we have considered the design where the plasma source is located in the gas inlet system (see Fig. \ref{ACI}).  However, there is another schematic when the process of plasma production can be achieved by a radial alternating electric field to propel a plasma discharge directly in the acceleration zone. From this technical point of view, such radial excitement is simpler than other methods for propelling plasma discharge. Herewith we can expect that the used configuration of external magnetic fields may increase the transfer processes of energy/momentum when the oscillation energy in the radial mode tends to be transferred to other degrees of freedom induced by the rotating plasma flows \cite{km17}-\cite{kys13}. However, in this case, one should take into account the coupling of ionization processes and nonlinear plasma oscillations \cite{k13,ksch}. This may be used with a transformation of energy/momentum radial oscillations directing into the axial direction by the plasma inhomogeneity with nonlinear coupling among the electron and ion flow components and oscillations \cite{kys10,ksy_2,k13}.

We would like to mention about the hypothetical possibility to consume an interplanetary medium as a fuel for driving plasma thrusters. This medium consists of plasma, neutral gas, dust (from nano- to micro-meter sized particles), and accelerated particles. The main component of the interplanetary medium is the solar wind, a supersonic plasma flow arising from the solar corona. The value of density for exploiting such a medium is extremely small but still finite. It means that under high velocities of the rocket, one may obtain a particle flux that can be ionized and used in a plasma accelerator. In order to fully understand this point, it will be necessary to analyze the properties of an interplanetary medium.  

\section{Conclusion}

The design of a plasma thruster that accelerates a plasma flow in the crossed magnetic fields [see Eq.(\ref{L_ccpt})] that uses a time-varying axial magnetic field $B_{z0}(t)\mathbf{e}_z$. This magnetic field provides an azimuthal acceleration of the plasma flow, and a permanent radial magnetic field $B_{r0}(t)\mathbf{e}_r$ provides the transformation of the azimuthal momentum into the axial momentum. Namely, the acceleration effect is caused by the azimuthal electric field which comes from the variation of an external magnetic flux [see Eq. (\ref{7_ac})]. Besides, it is possible to go beyond the approach of quasi-neutrality. These processes can lead to the generation of a local azimuthal time-dependent electric field which may accelerate the rotation of the electron and ion fluids in different directions. As previously mentioned, the interactions of these flows with the external magnetic field propel the axial acceleration of the flow. The presented relationships (\ref{L3_ccpt}) - (\ref{26a_ac}) define the range of allowable parameters of the device. In fact, the proposed design is a hybrid of induction accelerator and end-Hall thruster.

The authors are grateful to S.I. Krasheninnikov and V.A. Kurnaev for their interest in the work and helpful discussion. 

\end{document}